\documentclass[twocolumn,prb,a4paper]{revtex4}
\usepackage{epsfig}
\usepackage{dcolumn}

\begin{document}
\newcommand{\rar}{$\rightarrow$}
\newcommand{\lrar}{$\leftrightarrow$}

\newcommand{\beq}{\begin{equation}}
\newcommand{\eeq}{\end{equation}}
\newcommand{\bea}{\begin{eqnarray}}
\newcommand{\eea}{\end{eqnarray}}
\newcommand{\req}[1]{Eq. (\ref{Eq#1})}
\newcommand{\degree}{$^{\rm\circ} $}
\newcommand{\pcite}{\protect\cite}
\newcommand{\pref}{\protect\ref}
\newcommand{\rfg}[1]{Fig. \ref{F#1}}
\newcommand{\rtb}[1]{Table \ref{T#1}}

\title{Reversible B\lrar A Transitions in Single DNA Molecule Immersed in A
Water Drop}

\author{Alexey K. \surname{Mazur}}
\email{alexey@ibpc.fr}
\altaffiliation{FAX:+33[0]1.58.41.50.26}
\affiliation{Laboratoire de Biochimie Th\'eorique, CNRS UPR9080,
Institut de Biologie Physico-Chimique,
13, rue Pierre et Marie Curie, Paris,75005, France}
 

\begin{abstract}
Clarification of the detailed mechanisms involved in the DNA
polymorphism is an important challenge for computational molecular
biophysics. This paper reports about reversible A\lrar B transitions
in DNA observed {\em in silico} in a simulated titration experiment by
smooth variation of the size of a water drop around a double helical
solute. The estimated range of hydration numbers corresponding to the
B\lrar A transition roughly agrees with experimental data. The chain
length dependence was studied and it appeared that the transition to
A-form is hindered when the fragment becomes shorter than one helical
turn. Dynamics of the A\lrar B transition at low hydration is
cooperative and is driven mainly by medium range electrostatic
interactions of counterions sandwiched between phosphate strands in
the major groove. The correspondence of these computational
observations to common experimental conditions of A\lrar B transitions
is discussed. \end{abstract}

\maketitle

\section*{Introduction}

The transformation from B-DNA \cite{Watson:53} to dehydrated A-DNA
\cite{Franklin:53} in fibers was one of the first reversible
structural transitions observed in a biomolecule. These structures
represent right-handed helical duplexes with identical topologies and
hydrogen bonding, and yet they have strongly different shapes
\cite{Saenger:84,Calladine:92}. The B helix is long and narrow, with
stacked base pairs forming the core of the helix. In contrast, the
A-DNA structure is short and thick, with strongly inclined base pairs
wrapping around a 6 \AA\ solvent accessible cylindrical hole. The
B-form is the dominant biological conformation of DNA whereas the
A-form is considered as a high energy state that some sequences can
adopt temporarily during various biological functions
\cite{Ivanov:95}. Indeed, local A-DNA motives are rather common in
crystal structures of protein-DNA complexes \cite{Timsit:99,Lu:00}.
The A\lrar B polymorphism is currently considered as one
of the modes for governing protein-DNA interactions \cite{Ng:00}.

 A complete B\rar A transition in a long mixed sequence DNA
can be obtained {\em in vitro} in a few special conditions. In
crystalline \cite{Franklin:53,Leslie:83} and amorphous
\cite{Piskur:95} fibers as well as in films \cite{Tunis-Schneider:70},
it is induced by equilibrating the samples under reduced relative
humidity or with the concentration of organic solvents increased to
$\approx$80\% \cite{Zimmerman:83a}. Organic solvents can also provoke
B\rar A transition in an isolated DNA molecule in solution
\cite{Ivanov:83}. The transition is reversible and cooperative,
\cite{Ivanov:73,Ivanov:74} which means that it occurs concertedly in
an extended DNA fragment rather than randomly in separate nucleotides.
The characteristic length of such fragment, however, is is rather
small.  Originally, it was estimated as 20 base pairs
\cite{Ivanov:74}, but later this estimate was reduced to 10 base pairs
\cite{Ivanov:83}, that is approximately one helical turn. DNA
fragments shorter than this length generally do not go to the A-form
in solution \cite{Fairall:89,Galat:90}. For some sequences, however,
short DNA fragments could be crystallized in the A-form
\cite{Conner:82,NDB:}. Under physiological conditions {\em in vitro},
the A/B balance depends upon the environment, notably, the types of
solvent counterions \cite{Ivanov:73} and the temperature
\cite{Nishimura:86}. It is also sequence dependent, with some
sequences exhibiting in solution the features of both forms
\cite{Fairall:89,Wolk:89,Lindqvist:01}. The transition is
significantly facilitated in groups of consecutive guanines (G-tracts)
whereas in A-tracts it is more difficult
\cite{Leslie:83,Peticolas:88,Ivanov:96}. The A/B-philicity of
different base pair steps and triplets has been parametrized and used
successfully for predicting the properties of mixed sequences
\cite{Ivanov:95,Basham:95,Tolstorukov:01}.

Several different ideas concerning detailed molecular aspects and the
driving forces of A\lrar B transitions in DNA were
discussed in different years
\cite{Ivanov:73,Alden:79,Calladine:84,Calladine:92,Saenger:86,Hunter:93,Cheatham:97b,Cheatham:97c,Jayaram:98}.
Both the sugar phosphate backbone \cite{Saenger:86} and base pair
stacking \cite{Calladine:84} were proposed as one of the possible
actors in the transition, with water as an evident second partner.
Water bridges between certain DNA atoms may stabilize either B-DNA
\cite{Calladine:84} or A-DNA \cite{Saenger:86}, and the hydrophobic
effect possibly favors a B-DNA type of stacking
\cite{Ivanov:73,Hunter:93}. It has been suggested that the reduced
water activity is a universal factor that shifts the equilibrium
towards A-DNA \cite{Malenkov:75}, which was later replaced by a less
specific "economics of hydration" \cite{Saenger:86}. Although the
B\rar A transition induced by increased salt concentration has
been reported for poly-dG, the low water activity in high salt
generally does not induce B\rar A transitions
\cite{Ivanov:73,Zimmerman:80}. In any case, it is not clear how the
reduced amount of water around DNA affects its structure at the atomic
level \cite{Fuller:88}. Molecular dynamics simulations suggest, in
addition, that accumulation of metal cations in the major groove may
specifically stabilize A-DNA \cite{Cheatham:97b,Cheatham:97c}.

The cooperativity of the B\rar A transition is poorly
understood. Its physical origin was initially attributed to the sugar
puckering \cite{Ivanov:73,Ivanov:74}. The furanose ring conformation
is C3'-endo (North) in A-DNA (with the C3'-carbon puckered above the
furanose plane towards the nucleobase), and C2'-endo (South) in B-DNA
\cite{Saenger:84}. For a single nucleoside in vacuum, the C2'-endo
conformation is slightly lower in energy, with the barrier of 2-3
kcal/mole between the two states \cite{Foloppe:99}, and this is the
only significant local energy barrier between the A- and B-DNA
conformations. Until recently, it was assumed that the switch in the
sugar pucker is strictly linked with the base pair orientation. The
latter is restrained due to stacking interactions with the neighbors,
therefore, the sugar conformations can change only simultaneously in
at least several consecutive base pairs. These views, however, were
always questioned \cite{Calladine:84} and the recent X-ray and
spectroscopic studies really demonstrated that the North sugar puckers
are accessible without the B\rar A transition
\cite{Fairall:89,Wolk:89,Lindqvist:01}, and even with all sugars in
the C3'-endo conformation the overall B-DNA structure can persist
\cite{Ng:00}. It appears, therefore, that the sugar conformations are
only weakly restrained by the stacking and the observed cooperativity
can hardly be due to the sugar pseudorotation barrier. At the same
time, the ensemble of the currently available X-ray DNA structures
agrees with the cooperative character of the B\rar A
transition. Although some A-DNA structures have B-like features near
the termini \cite{Malinina:99}, and there exist a series of structures
classified as A/B transition intermediates \cite{Vargason:01} the A-
and B-forms generally do not mix in the same DNA fragment. The two
forms may even coexist in a crystal, but in different samples of the
same fragment \cite{Doucet:89}.

Many aspect of the A/B polymorphism were earlier studied by theoretical
methods. Its thermodynamics is well described by an analytical theory
based upon the Ising model \cite{Ivanov:74}.  Reduced
\cite{Calladine:84,Hunter:93,Mazur:89} and all atom \cite{Ivanov:79}
models in continuous dielectric were used for studying its mechanism
and sequence dependence. On the other hand, different analytical
theories combined with simplified DNA representation were used for
modeling solvent effects \cite{Jayaram:96}. Steady progress in MD
simulations of DNA since the pioneering calculations by Levitt
\cite{Levitt:83b} recently made possible detailed modeling of A/B
transitions in realistic environment including explicit water and
metal counterions
\cite{Cheatham:97b,Cheatham:97c,Jayaram:98,Cheatham:96,Yang:96,Cieplak:97,Sprous:98,Langley:98,Soliva:99}.
They shed new light upon the putative molecular mechanisms involved in
the DNA A/B polymorphism despite limited time of MD trajectories and
some forcefield artifacts \cite{Feig:98}.

In the present study I try to get new insight in this problem by using
a different MD approach. It has recently been shown that the
particle-mesh Ewald (PME) algorithm allows one to follow the dynamics
of a DNA double helix put in a small salty water drop
\cite{Mzjacs:02}. An evident question one asks, what happens if the
size of the drop is reduced? By analogy with drying fiber samples it
should be expected that DNA would go from B to A-form. A series of
simulations described here confirms that this is really the case.
Although these conditions have never been used in experiments our
simulations give the best currently possible prediction of the
properties of such systems. Moreover, the transition is reversible in
the time scale accessible in the calculations, that is both B\rar A
and A\rar B transitions can be observed in the nanosecond time range.
This allows one to test different hypotheses concerning the factors
governing the transition. The results suggest that the A\lrar B
transitions in an isolated water drop are driven by medium range
electrostatic interactions between DNA backbone and counterions
accumulated in the major groove. Moreover, several evidences suggest
that A\lrar B transitions in other conditions are actually governed by
the same mechanism.

\section*{Methods}

In the series MD simulations presented here the Dickerson-Drew double
helical dodecamer (CGCGAATTCGCG \cite{Wing:80}) and its derivatives
are used as examples. Similar results were also observed for several
other test sequences. This particular DNA fragment is preferred
because it is neither A-philic or B-philic and because its dynamic
structure with Cornell et al. force field \cite{Cornell:95} is rather
close to experimental data \cite{Cieplak:97,Young:97b,Duan:97}. The
simulation protocols were similar to the earlier water drop
simulations \cite{Mzjacs:02}.  We use the internal coordinate
molecular dynamics (ICMD) method \cite{Mzjcc:97,Mzbook:01} adapted for
DNA \cite{Mzjacs:98,Mzjchp:99} with the time step of 0.01 ps. The
electrostatic interactions are evaluated with the shifted Coulomb law
by using the SPME method \cite{Essmann:95,Mzjacs:02} and the long
range cut-off of 50 \AA\, which was always larger than the system
size.  The Van-der-Waals and SPME direct sum interactions were
truncated at 9 \AA\ with the value of Ewald parameter $\beta\approx
0.35$.

The initial state for B\rar A transitions was prepared as before
\cite{Mzjacs:02} with the canonical B-DNA \cite{Arnott:72} used a
standard conformation. The DNA molecule was first immersed in a large
rectangular TIP3P \cite{Jorgensen:83} water box of and next external
solvent molecules were removed by using a spherical distance cut-off
from DNA atoms. The cut-off radius was adjusted to obtain the desired
number of water molecules remaining. The drop was neutralized by
randomly adding the necessary number of Na$^+$ ions. Simulations of
A\rar B transitions started from the final structure of a B\rar A
transition obtained with the lowest hydration.

Every system was energy minimized first with the solute held rigid and
then with all degrees of freedom. Dynamics were initiated with the
Maxwell distribution of generalized momenta at 250K and equilibrated
at this temperature during several picoseconds.  Production
trajectories were computed with the temperature bound to 300 K by the
Berendsen algorithm \cite{Berendsen:84} with a relaxation time of 10
ps. For better comparison with earlier simulations of A\lrar B
transitions, the original Cornell et al. force filed \cite{Cornell:95}
was used.  Several test simulations showed, however, that
qualitatively similar results are obtained with the subsequent
modifications of the AMBER force filed \cite{Cheatham:99,Wang:00}.
Duration of production runs varied from 5 to 25 ns depending upon the
character of dynamics observed. The conformations were saved with a
2.5 ps interval.  Programs Curves, \cite{Curves:} XmMol \cite{XmMol:}
and Mathematica by Wolfram Research Inc. were employed in the data
processing.

\section*{Results and Discussion}

\begin{figure}
\centerline{\psfig{figure=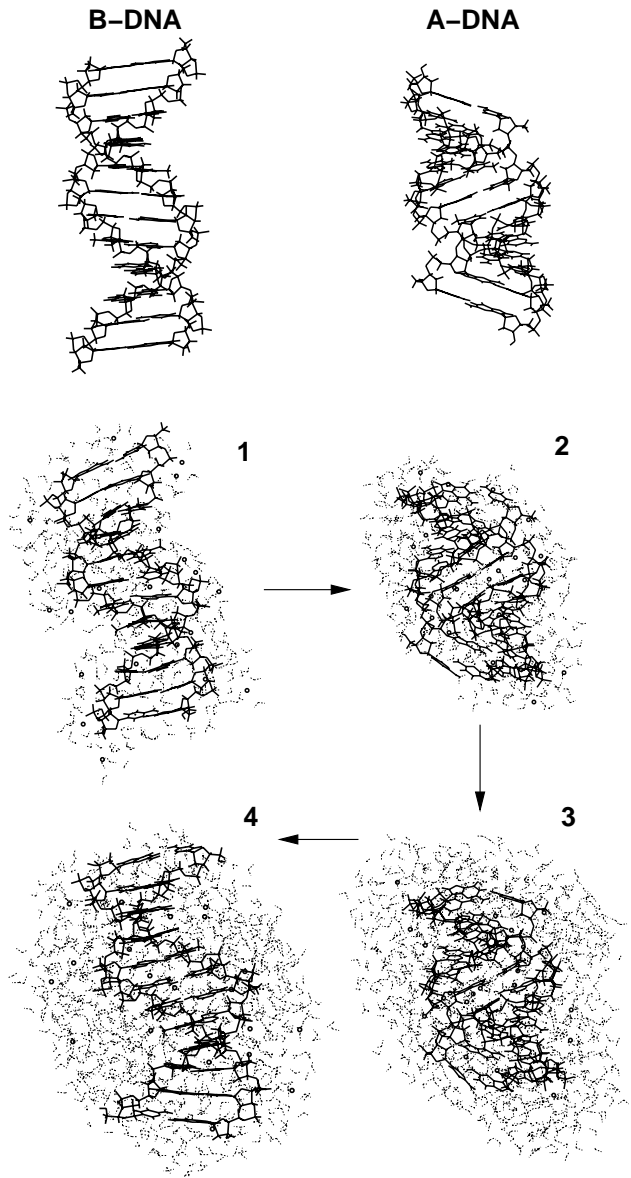,width=9cm,angle=0.}}
\caption{\label{Fsnaps}
An example of a reversible B\rar A transition.
Water molecules and Na$^+$ ion positions are shown by broken lines
and large dots, respectively.
}\end{figure}

\begin{table}[t]\caption{\label{Tenpa}
Some structural parameters of DNA conformations after either A\rar B
or B\rar A transition obtained in dynamics. The drop size is given as
the number of water molecules, with their number per nucleotide in
parentheses. The rmsd from the canonical A and B-DNA and sequence
averaged helical parameters\pcite{Curves:} are computed for
conformations averaged over the last nanosecond of dynamics.  All
distances are in angstr{\"o}ms and angles in degrees.  The approximate
time of transition (in nanoseconds) corresponds to the end of the
sugar switch phase}
\begin{ruledtabular}
\begin{tabular}[t]{|ccccccc|}
Size
   & Transition
              & RMSD-A
                     & RMSD-B
                            & Xdisp
                                  & Inc
                                      & Time
\\
\hline
 400(16.7) & B\rar A &  2.1 & 6.5  & -4.2 & 23.0 &  1.5  \\
 500(20.8) & B\rar A &  1.8 & 6.4  & -4.0 & 21.2 & 18.0  \\
 600(25.0) & B\rar A &  1.8 & 6.1  & -4.0 & 20.2 & 20.0  \\
 800(33.3) & A\rar B &  3.3 & 3.4  & -2.9 &  9.9 &  5.0  \\
1000(41.7) & A\rar B &  3.3 & 3.4  & -3.2 &  9.4 &  1.0  \\
2000(83.3) & A\rar B &  3.5 & 3.8  & -3.4 &  9.9 &  1.5  \\
\end{tabular}
\end{ruledtabular}
\end{table}

\begin{figure*}
\centerline{\psfig{figure=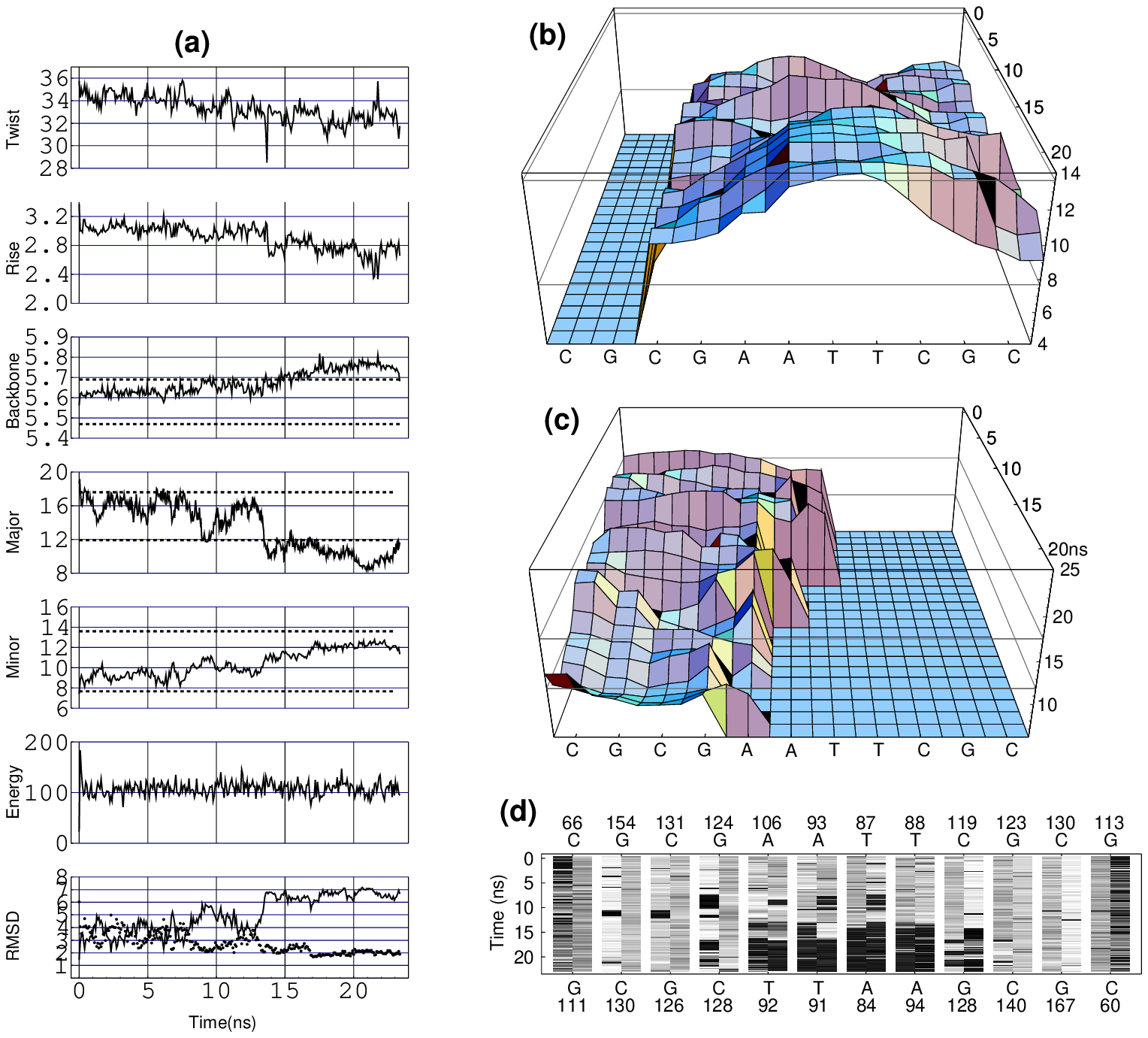,width=16cm,angle=0.}}
\caption{\label{FktBA}
A representative example of dynamics of a B\rar A transition.\\ (a)
Time traces of some important parameters. From top to bottom: the
average twist and rise; the average backbone length computed as the
distance between the centers of consecutive sugar rings; the average
width of the major and minor grooves; the total energy; the rmsd from
the canonical A-DNA (dotted line) and B-DNA (solid line).  The
horizontal broken lines in some frames mark the canonical B and A-DNA
values.\\ (b) The time evolution of the minor groove profile. The
surface is formed by 23 minor groove profiles equally spaced in time
and each averaged over a 75 ps interval. The profile on the front face
corresponds to the final DNA conformation. The groove width is
evaluated by perpendicular cross-sections of space traces of C5' atoms
\pcite{Mzjmb:99}, with the measured distances assigned to the sequence
of the upper ("Watson") strand. The empty zone on the left corresponds
to the region where the groove could not be measured. The width is
given in angstr\"oms and the corresponding canonical A- and B-DNA
levels of are marked in all plates by the thin straight lines.\\ (c) The
time evolution of the major groove profile. The surface is constructed
similarly to plate (b). The empty zone on the right corresponds to
the region where the groove could not be measured.\\ (d) Dynamics of
sugar pucker pseudorotation. Only the pucker phase values from 0 to
180 were considered. Sugars with phases beyond this interval were
assigned the closest of the two boundary values. The minimal and the
maximal phases obtained were assigned the white and the black colors,
respectively, with intermediate values mapped linearly to the gray
scale levels.  Each base pair step is characterized by a column
consisting of two sub-columns, with the left sub-columns referring to
the sequence written above in 5'-3' direction from left to right with
the time averaged phases given on top. The right sub-columns refer to
the complementary sequence shown below together with the corresponding
time averaged phases.}\end{figure*}

An example of the molecular transformation we study is shown in
\rfg{snaps}. The top row exhibits the canonical B- and A-DNA
structures. The process begins with plate 1 where we see a snapshot of
the first saved state during a production run started from the
canonical B-DNA conformation. This system involves a dodecamer DNA
fragment with 22 Na$^+$ and 400 water molecules. It is seen that after
the equilibration some ions have already entered the major groove
while a thin water shell covers the surface of the DNA molecule. One
may notice also a small narrowing in the middle of the minor groove
which is a well-known distinctive feature of the crystal structure of
this dodecamer. This initial trend is only temporary, however. The
number of water molecules in the system is about 17 per nucleotide,
which is less than the limiting hydration number for B-DNA
\cite{Saenger:84}.  In dynamics, water tries to form a smooth
spherical drop around the DNA fragment and this is achieved owing to a
B\rar A transition. Plate 2 shows a snapshot after 2 ns of dynamics.
It is seen that the DNA molecule went to the characteristic A-form
conformation with a wide minor and narrow major grooves. Note that a
substantial part of the Na$^+$ ions is sandwiched in the major groove
between the two opposite sugar-phosphate strands.  The more compact
A-form structure allows water to form a nearly spherical drop that
covers the DNA fragment.  After the transition the structure remains
stable during the time accessible in calculations.

The final state of the previous trajectory was used as a starting
point for the second run to obtain an inverse A\rar B transition. For
this purpose the system was surrounded by 600 additional water
molecules to increase their number to 1000. Plate 3 in \rfg{snaps}
displays a snapshot from the beginning of the second production run.
Now the number of water molecules is about 42 per nucleotide, which
corresponds to B-DNA hydration numbers.  The last plate in
Fig. \ref{Fsnaps} shows a snapshot after 2 ns of dynamics. As we see,
the DNA molecule really went back from A- to B-DNA conformation and
the narrowing in the middle of the minor groove re-appeared.

The sequence of transitions shown in \rfg{snaps} suggests that there
exists a critical level of hydration characterized by an equal
probability of A and B-forms, if the transition is cooperative, or by
some structures intermediate between the two.  In order to get a rough
estimate of this midway point, a series of MD simulations was carried
out with the size of the water drop systematically varied. The results
obtained are summarized in \rtb{enpa}. The corresponding trajectories
were started from standard A-DNA and B-DNA states and continued as
long as necessary to reach either A\rar B or B\rar A transition. For
intermediate hydration values, notably, for 21 and 25 water molecules
per nucleotide, very long trajectories were necessary. In all cases a
transition in one direction only was obtained. For instance, with 25
water molecules per nucleotide an alternative trajectory starting from
A-DNA was continued to 10 ns, but no transition occurred. The results
summarized in \rtb{enpa} suggest that the hydration of 25 water
molecules per nucleotide can serve as a rough estimate of the
transition point.

As seen in \rfg{snaps}, the A-DNA structure obtained in dynamics is
more compact than the canonical conformation. The major groove width
in the structure averaged over the last nanosecond is about 2 \AA\
below the canonical value. The final A-DNA structures obtained for
higher hydration levels had slightly wider major grooves. Note that,
in \rtb{enpa}, their rmsd from the canonical A-DNA is lower. In all
cases, however, structures with the major groove width similar to that
in the canonical A-DNA were observed only temporarily in intermediate
transition phases, and they seemed to be unstable with
respect to the more compact conformation shown in plate 2 of
\rfg{snaps}.  Visual inspection reveals that in such structures the
phosphate groups opposed across the major groove are similarly close
or even closer to each other than the neighboring phosphates in the
same strand.  They are linked by multiple water bridges and at least
one such bridge is usually replaced by a Na$^+$ ion slightly shifted
either inside or outside the helix. With intermediate hydration
numbers, however, this pattern was less clear, with both Na$^+$ and
the opposite phosphate groups often hydrated separately.

One of the trajectories mentioned in \rtb{enpa} is detailed in
\rfg{ktBA}. This simulation started from the canonical B-DNA with a
water shell of 500 molecules (20.8 per nucleotide). Plate (a) exhibits
time plots of several structural parameters. Three of them, namely,
the backbone length and the average width of the two grooves deserve
preliminary comments. The backbone length is computed as the average
distance between the centers of consecutive sugar rings. Note that its
value is larger in A-DNA than in B-DNA because, in a B\rar A
transition, the sugars are moved to the outer surface of the double
helix, with the helical diameter of the sugar trace increased by
nearly 6 \AA. The  widths of both grooves are evaluated by
perpendicular cross-sections of spatial traces of C5' atoms
\pcite{Mzjmb:99}. Near the helical termini, however, the grooves are
not measurable and, in fact, the groove lengths vary in dynamics.  The
last feature is very clear in the dynamics of the major groove profile
shown in plate (c). The average widths were computed for the zones
where the corresponding groove was defined in each given structure.

The B\rar A transition seen in \rfg{ktBA} is relatively slow, with the
final structure established only after 18 ns. An early B\rar A
transition occurred close to one of the termini at around 10 ns.
However, it was incomplete and inverted two nanoseconds later.
According to the two rmsd traces, the final transition started at 13
ns. Until that the dynamics sampled midway conformations between A-
and B-forms, which is not that far from normal B-DNA dynamics with the
same potentials \cite{Cieplak:97,Young:97b,Duan:97}. It may be seen
that the minor groove profile periodically switched from narrowing to
widening in the middle. At about 13 ns the two rmsd traces drastically
moved close to their final levels. Simultaneously the major groove
width fell by almost 10 \AA\ and went below the A-DNA level.  At that
time the molecule already looked very much like A-DNA, with the
instantaneous rmsd-A values around 2 \AA, but almost all sugars still
had their puckers in the South region.  The final phase of the
transition is slower and during this phase the sugars in the middle
AATT tetraplet switch to the North. In this middle zone the transition
to the A-form is complete whereas the three terminal base pairs remain
in an intermediate conformation with South sugar puckers.

Although the structural parameters shown in plate (a) differ by the
amplitudes of fluctuations they all reveal a slow overall drift
suggesting a downhill motion of the system on a free energy surface.
In contrast, the total energy remains stable throughout the trajectory
except the initial phase. This means that, already during heating, the
system finds a balance of internal interactions maintained afterwards,
and, consequently, the observed steady downhill character of the
transition has an entropic origin. The precise nature this entropic
force is not clear. The backward A\rar B transitions have
qualitatively similar kinetic suggesting that this feature is not due
to the properties of the A and B-DNA forms, but results from the
specific setup of simulations.

In the backward A\rar B transitions, the sequence of events is
inverted, that is, the sugar puckers quickly switch to the North
whereas the rmsd's and the groove widths need sometimes many
nanoseconds to reach stationary B-DNA values. Such kinetics does not
depend upon the amount of water added and it is slower than in earlier
reported AMBER simulations \cite{Cheatham:96,Sprous:98}. The apparent
reason is that during the B\rar A transition a large number of metal
ions enter the major groove when it is still wide. A high ion
concentration in the major groove is one of the factors that
supposedly stabilize the A-DNA conformation
\cite{Cheatham:97b,Cheatham:97c}. In contrast, when the trajectory is to
be started from the canonical A-DNA, the major groove is narrow from
the beginning, and very long equilibration period should be necessary
for the same number of ions to reach optimal positions.

\begin{figure}
\centerline{\psfig{figure=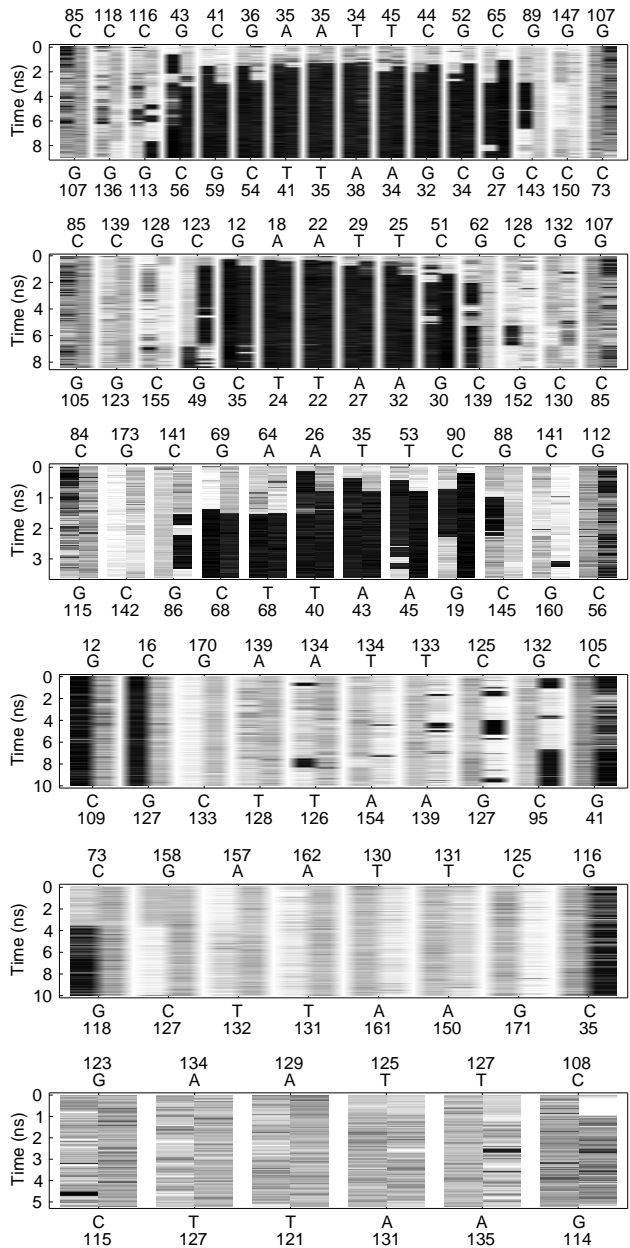,width=9cm,angle=0.}}
\caption{\label{Ffrag} Dynamics of B\rar A transitions in DNA
fragments obtained by elongation and truncation of the initial
sequence. For details, see the legend to plate (d) of \rfg{ktBA}.
}\end{figure}

The possibility of B\lrar A transitions in longer and shorter DNA
fragments has been checked in a series of simulations summarized in
\rfg{frag}. The foregoing results suggested that the B\rar A
transitions in a water drop are mainly driven by electrostatic
interactions between counterions and backbone phosphates across the
major groove.  A characteristic A-form major groove can be formed
starting from eight base pairs and, as seen in \rfg{ktBA}c, its
length in a dodecamer A-DNA is between for and five nucleotides.  One
could expect, therefore, that B\rar A transitions would progressively
become more difficult as the DNA is shortened.  All trajectories shown
in \rfg{frag} started from B-DNA with the lowest degree of hydration
used in \rtb{enpa}. The number of Na$^+$ and the size of the water
drop were changed according to the DNA chain length. As expected,
relatively rapid B\rar A transitions are observed in the original
dodecamer and its longer derivatives. In contrast, the transition did
not occur in shorter fragments. The hexamer always remained in a
typical B-DNA conformation with high rise and negative inclination.
The octamer and decamer fragments exhibited dynamics somewhat similar
to that of the dodecamer, with accumulation of Na$^+$ ions in the
major groove and noticeable compression of the DNA structure, however,
the B\rar A transition did not occur. Moreover, one of the terminal
base pairs in the decamer was broken, with water bridges formed
between the two nucleotides, which apparently helped to move closer
the opposite phosphate strands in the major groove. This effect was
reproduced in a repeated run with hydrogen bonds in terminal base
pairs enforced by weak additional restraints. The last observation
suggested that the three shorter fragments in \rfg{frag} would not go
to A-form even in much longer dynamics.

Based upon the described qualitative character of DNA dynamics, one
can consider two main physical factors as the possible driving forces
of the B\lrar A transition under low hydration. First, water tends to
reduce its contact surface with vacuum and forms a spherical drop
forcing the DNA fragment to shrink. This force can be considered as
surface tension or as a particular case of the hydrophobic effect,
with vacuum in place of a nonpolar solvent. The second factor is the
increased concentration of Na$^+$ ions, which reduces the water
activity. These two alternatives were checked in additional
simulations under modified conditions. To check separately the role of
the surface tension, the counterion effects were suppressed by
discharging the DNA fragment and next running dynamics without
counterions. The partial charges of OP atoms were reduced by by 0.5 eu
and two long trajectories were computed starting from the canonical
B-DNA structure under the lowest and the highest hydration,
respectively. The DNA structures sampled in both these simulations
turned out to be similar. These were slightly underwound B-DNA
conformations with the average twist around 30\degree. Under the low
hydration the duplex looked slightly compressed, but no signs of a
B\rar A transition could be noticed.

\begin{figure*}
\centerline{\psfig{figure=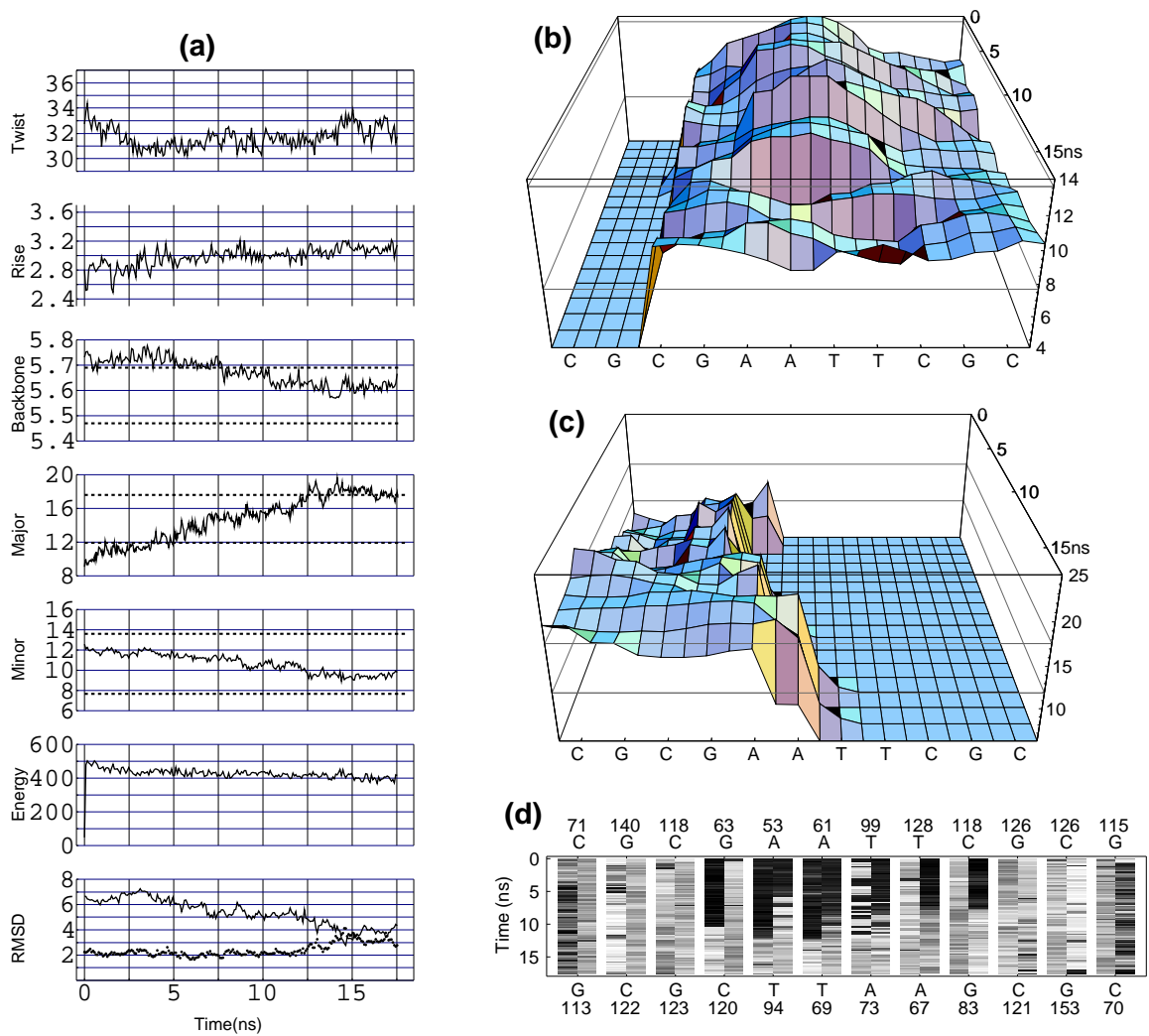,width=16cm,angle=0.}}
\caption{\label{FktAB}
A dynamics of a B\rar A transition in a 600 water molecule drop with
increased NaCl concentration. For details, see the legend to
\rfg{ktBA}.
}\end{figure*}

The effect of the high counterion concentration upon the water
activity was separately checked in two simulations with added NaCl.
Earlier it was reported that high NaCl does not cause B\rar A
transitions in conventional simulations with periodical boundaries
\cite{Cheatham:97c}. To make a more stringent test, here the NaCl
concentration was increased under the midway hydration as estimated in
\rtb{enpa}. It was hoped that under such conditions an increased salt
concentration can produce a clear qualitative effect upon the dynamics.
The initial A and B-DNA states in a water shell of 600 molecules were
modified by replacing 66 randomly chosen water molecules with 33
Na$^+$ and 33 Cl$^-$ ions. After that, 66 water molecules were added
to maintain the earlier hydration level.  The effective concentration
of Na$^+$ was thus raised from 2M to 5M.  The two systems were
re-equilibrated with the standard protocol and long trajectories were
computed to obtain either A\rar B or B\rar A transition. In contrast
to the tests in \rtb{enpa}, here an A\rar B transition was obtained
after about 13 ns dynamics as shown in \rfg{ktAB}. The dynamics
exhibited in this figure is typical of the A\rar B transitions
described above although its kinetics are much slower than those in
\rtb{enpa}. The second trajectory was similarly long, but only B-DNA
structures were sampled.

In the animations of these trajectories, progressive crystallization
of NaCl was seen, with large blocks of typical cubic NaCl lattice
found in the major DNA groove in the end of both trajectories.
Addition of NaCl is supposed to stimulate A\rar B transitions in fiber
DNA crystals\cite{Cooper:66}. Moreover, NaCl crystallization occurs in
DNA fibers at 75\% relative humidity if the salt content is above 6\%
of the DNA weight\cite{Cooper:66}, which gives 4.8M effective
concentration of NaCl according to the reported water/DNA adsorption
curves\cite{Falk:62}. One may note, therefore, that the overall effect
of the added salt in our simulations qualitatively agrees with
experiments. The observed shift towards the B-form may be caused by
the NaCl crystallization either directly, by forcing the major groove
to open, or indirectly, by pushing water out from the major groove
together with some free Na$^+$ ions. In any case, however, these
results suggest that it is not the water activity reduced by ions that
causes the B\rar A transitions in small drops.

The last simulations suggest that B\rar A transitions are induced by a
united effect of the limited water shell and counterions, and not by
one of these factors considered separately. A simple and attractive
mechanism that accounts for all the above results is as follows. The
B\rar A transition is caused by direct electrostatic interactions
between mobile metal cations and phosphate groups in the opening of
the major DNA groove. When the local cation concentration exceeds a
certain critical value, the inter-phosphate repulsion is inverted to
attraction. A similar idea is long discussed in the context of DNA
bending\cite{Mirzabekov:79,Levene:86,Rouzina:98} and it was earlier
proposed by Cheatham and Kollman as one of the possible mechanisms
responsible for stability of the A-form
\cite{Cheatham:97b,Cheatham:97c}. The strong inter-phosphate
attraction immediately results in local DNA bending and, eventually,
in a B\rar A transition at the opposite side of the double helix. When
the length of the major groove exceeds one helical turn the apparent
bend becomes uniform shrinking. The role of the limited water shell
consist in pushing the counterion cloud surrounding DNA inside the
diameter of the double helix. The counterions avoid loosing their
hydration shells and are forced to come closer to DNA. The surface
tension at the drop boundary plays a certain role, but it is the
reduced overall space available for counterions that seems more
important. This simple model explains well our results, and below we
consider why and how it may work in other conditions as well.

\subsection*{Concluding Discussion}

Clarification of the detailed mechanisms involved in the DNA
polymorphism is an important challenge for rapidly progressing
computational methods in molecular biophysics\cite{Cheatham:00}. Here,
for the first time, reversible B\lrar A transitions were obtained {\em
in silico} in a simulated titration experiment by smooth variation of
water content. The environment conditions used were not earlier
encountered in experiments, nevertheless, the qualitative features
found seem credible and likely to exist in analogous experimental
conditions. First, the computed pattern of DNA transformations is
physically sensible and agrees with known experimental trends,
notably, as regards the principal effect of the amount of bound water.
Second, these calculations give the most accurate currently possible
prediction of the true properties of DNA fragments in small water
drops. Really, the Cornel at al. force field \cite{Cornell:95} was
earlier shown to produce DNA dynamics in good agreement with
experiments \cite{Cheatham:00}. Its parameters were fitted with
small molecule data only, and they were never specifically adjusted
for periodical boundary conditions. In our case, no artificial
interactions are introduced, while those between periodical images are
eliminated \cite{Mzjacs:02}. There are all reasons to believe,
therefore, that our simulations reproduce correctly the dynamics of
DNA transformations in such unusual conditions.

Qualitatively, the results obtained here agree with earlier studies.
Cheatham and Kollman were the first to obtain an A\rar B
transformation in pure water, and stable A-DNA dynamics in 85\% EtOH
\cite{Cheatham:96,Cheatham:97b}. Visual inspection of the reported
A-DNA snapshots \cite{Cheatham:97a} shows that our structures are
quite similar to those in EtOH as regards the width of the major
groove, preferred positions of Na$^+$ ions, and a B-like character of
the DNA termini, which gives the molecule a strong apparent bend to
the major groove. Moreover, the reported stabilization of A-DNA in
EtOH\cite{Cheatham:97b} required that the molecule was first covered
by water and then by EtOH as an outer shell. The necessity of such
double phase hydration was confirmed by Sprous et al.
\cite{Sprous:98}.  Although water and EtOH could mix in dynamics,
it seems evident that, in order to be stable in simulations, the
A-form should be contained in a small water shell.  Cheatham and
Kollman showed also that a B\rar A transition can be provoked by
placing polycationic ligands to the major groove between the opposite
phosphate strands \cite{Cheatham:97c}. Our results confirm that the
accumulation of positively charged counterions in the major groove of
B-DNA is, perhaps, the major driving force of this transition.

Although one cannot exclude that the A\lrar B transitions in different
conditions are due to radically different mechanisms, it is attractive
to consider the possibility that the electrostatic model outlined above
in fact operates in all such cases. This model is obviously
transferable from water drop to DNA in fibers under different
humidity.  One can argue, tough, that the computed A-DNA
systematically differs from the canonical and single crystal A-DNA
fragments by a very narrow major groove. However, the major groove in
the A-DNA conformations available in NDB\cite{NDB:} varies so strongly
that this parameter can hardly be considered as a well defined
attribute of the A-form. It is possible also that the parameterization
of Na$^+$ ions in the Cornell et al. force field\cite{Cornell:95}
underestimates its binding to water with respect to negatively charged
atoms. In our dynamics, correct canonical major groove width was
observed during transitions as long as Na$^+$ ions kept their first
hydration shells and avoided direct contacts with the phosphate
groups.

Another defect of the computed A-DNA consists in the strong
B-philicity of the termini accompanied by bending to the major groove.
These two features are mutually related because it is long known that
the boundaries between A and B-forms are bent\cite{Selsing:79}. This
defect, however, is only apparent because it qualitatively agrees with
experimental data. The B-philicity of DNA ends has to be assumed when
A\lrar B transitions in solution are considered\cite{Minchenkova:86}.
Moreover, many single crystal A-DNA structures appear smoothly bend
to the major groove when analyzed with the Curves
program\cite{Curves:}, with this trend being particularly evident for
complete helical turns \cite{Malinina:99,Verdaguer:91,Bingman:92}. In
our simulations, these features are just accentuated due to the
excessively narrow major groove, which requires stronger bending, and
also because our sequences are not A-philic as in all single
crystal A-DNA fragments.

The solution A\lrar B transitions caused by non-polar organic solvents
can also be interpreted in terms of the above electrostatic model.
For this purpose it is sufficient to imagine what would happen if our
water drops were put in EtOH, for instance. The water activity inside
the drop is strongly reduced by DNA itself and its counterion shell.
In infinite water, the activity coefficient grows with the distance
from the double helix and becomes close to one outside the counterion
cloud, that is at around 10 \AA\ from DNA. EtOH is easily mixable with
normal water, but not with the low activity water near DNA. This
argument can be illustrated by the above discussed crystallization of
NaCl in the DNA environment at concentrations much below its standard
solubility limit. The water molecules would leave the drop and go to
EtOH until the outer water activity becomes equal to that inside the
drop.  In other words, the water activity in the added nonpolar phase
determines the size of the drop and, consequently, the state of the
A\lrar B equilibrium. That is why B\lrar A transition curves in
different solvents converge when water activity is used as
parameter\cite{Malenkov:75}. More polar solvents like MetOH and
ethylene glycol do not cause B\rar A transitions possibly because they
solvate counterions rather than push them inside the double helix.

Both the apparent B-philicity of the DNA ends and the chain length
dependence of the B\rar A transitions observed in our simulations
agree with experiments in
solution\cite{Minchenkova:86,Fairall:89,Galat:90}.  Another salient
feature nicely reproduced is the cooperative character of the
transition, with the minimal length of the cooperative fragment of
around one helical turn. This cooperativity is not caused by the sugar
pseudorotation barrier because during B\rar A transitions the sugars
switch last. Instead it is conditioned by the A-DNA geometry and the
character of the electrostatic interactions in the major groove. In
the A-DNA major groove the phosphate of the nth base pair is positioned
against that of the (n-8)th. These eight base pairs represent the
minimum DNA length that is to be mechanically involved if the
transition is caused by a force applied between these groups. The
positive energy of this mechanical deformation should be compensated
by the negative electrostatic energy of solvent cations sandwiched
between the phosphates. One step of such "electrostatic sandwich" is
probably insufficient, and octamer DNA should not generally go to the
A-form. However, already in a nanomer the sandwich length is increased
by one while the deformation energy is increased only by a fraction of
the initial gap. Therefore, the energetic balance is progressively
shifted in favor of A-form as the chain length is increased. In long
DNA, the boundary between the A and B-forms has higher energy because
the energy gap of the initial octamer remains effectively shared
between the two opposite boundaries. The foregoing qualitative pattern
exactly corresponds to the one dimensional Ising model of cooperative
transitions\cite{Ivanov:74}.

\bibliography{onecol}
\bibliographystyle{jcp}

\end{document}